\documentclass[12pt,preprint]{aastex}
\usepackage{natbib,graphicx}

\shorttitle{Black Hole Spins}
\shortauthors{Daly}
  
\begin{document}
 
\title{Bounds on Black Hole Spins}
\author{Ruth A. Daly\altaffilmark{~}}
\affil{Department of Physics, Penn State University, Berks Campus, P. O. 
Box 7009, Reading, PA 19610}
\email{rdaly@psu.edu}

\begin{abstract}
Beam powers and black hole masses of 48 extended 
radio sources are combined to 
obtain lower bounds on the spins and magnetic field strengths 
of supermassive black holes.  This is done
in the context of the models of Blandford \& Znajek (1977) (the 'BZ' model) 
and Meier (1999); a parameterization for bounds in the context of other models
is suggested. The bounds obtained 
for very powerful classical double radio sources 
in the BZ model are consistent with 
black hole spins of order unity for sources at high redshift. 
The black hole spins are largest for the highest
redshift sources and decrease for sources at lower redshift; the
sources studied have redshifts between zero 
and two. Lower power radio sources associated with central
dominant galaxies may have black hole spins that are significantly
less than one.   
Combining this analysis with other results suggests that the maximum values of 
black hole spin associated with powerful radio galaxies 
decline from values of order unity at a redshift of 2 to
values of order 0.7 at a redshift of zero, falling roughly as $\sqrt(1+z)$, 
while lower power radio sources have spin values that range from about
0.1 to 0.8. These black hole spin values decrease if the data are considered
in the context of the Meier model rather than the BZ model.

\end{abstract}


\section{INTRODUCTION}

Quasars and other types of AGN activity are believed to be powered by 
supermassive black holes.  The AGN activity may produce 
highly collimated outflows from the immediate vicinity of the 
black hole (e.g. Rees 1984; Blandford 1990), which can  
power a large-scale radio source.  
Two defining properties of astrophysical black 
holes that can be measured in principle are the black hole mass and spin.
A significant
amount of progress has been made toward measuring the masses of
supermassive black holes (e.g. Kormendy \& Richstone 1995; 
Magorrian et al. 1998; Ferrarese \& Merritt 2000; Gebhardt et al. 2000; 
Ferrarese \& Ford 2005), though measuring the spin 
has been more challenging.

General theoretical studies suggest that 
the merger and accretion history of a supermassive black hole is
encoded in the spin of the hole.  For example, successive mergers
are likely to produce black holes that are spinning at a moderate
rate ($j \sim 0.7$), while powerful accretion events are 
likely to produce rapidly rotating black holes (Berti \& Volonteri 2008). 
Alternatively, most black holes may grow due to many short-lived, 
uncorrelated accretion episodes, which would lead to lower spin
values (King \& Pringle 2006, 2007).
Numerous uncorrelated accretion 
episodes tend to cause the spin of the hole to decrease over  
time with substantial fluctuations in spin caused by each episode
(King, Pringle, \& Hofmann 2008). 

At present, only a few 
observations allow black hole spins to be studied directly. 
Observations of Seyfert galaxies suggest rapidly rotating  
black holes in these systems (Wilms et al. 2001; 
Fabian et al. 2002).  A large spin is indicated by 
observations of the Galactic center black
hole (Genzel et al. 2003; Aschenbach et al. 2004). And, 
X-ray observations of active galaxies suggest
rapidly rotating black holes in these systems (Crummy et al. 2006). 
Outflows from AGN that produce extended radio sources 
allow a lower bound to be placed on the spin of the black hole 
that powers the outflow.  
Assuming only that the outflow is powered by the spin energy of the
black hole, Daly (2009) showed that 
the outflow energy and the black hole mass may be
combined to obtain a lower limit on the spin of the black hole.   
For a sample of 19 very powerful classical double sources, the 
lower bound was about the same for each source in the sample, 
and indicated $j_{min} \approx 0.12 \pm 0.01$.

Here, beam powers and black hole masses of radio sources 
are combined to 
study black hole spins in the contexts of 
the models of Blandford \& Znajek (1977), Meier (1999),
and models with similar functional forms. It 
is shown that the spin and magnetic field strength of a supermassive
black hole can be rather tightly constrained 
in the contexts of these models. 
The method is described in section 2.  It is applied to 
two samples of sources, and the results are presented in section 3, 
discussed in section 4, and summarized in section 5. A standard cosmological
model with $H_0 = 70$ km/s/Mpc, $\Omega_m = 0.3$, $\Omega_{\Lambda}=0.7$,
and zero space curvature is assumed throughout.

\section{THE METHOD}
\label{method}

The well-known model
of Blandford \& Znajek (1977), referred to as the 'BZ' model, 
and other models to power highly collimated outflows from AGN
are considered here.    
In the BZ model, the beam power $L_j$
that will be produced  is related to the dimensionless black hole spin 
$j \equiv Sc/(GM^2)$,  the black hole horizon 
$r_H = 2GM/c^2$, 
and the poloidal magnetic field strength $B_{p0}$ at the horizon,
where $S$ is the black hole spin angular momentum,  
M is the black hole mass, 
G is Newton's constant, and c is the speed of light 
(e.g. Macdonald \& Thorne 1982; Thorne, Price, \& Macdonald 1986) 
\begin{equation}
L_{j}(BZ) = (j^2 B_{p0}^2 r_H^2 \omega_F^2 c)/32 \approx 2 \times 10^{43}
j^2 M_8^2 B_4^2\hbox{ erg/s}.
\end{equation}
Here $B_4$ is the value of $B_{p0}$ in units of $10^4$ G,  
$M_8$ is the black hole mass in units of $10^8 M_{\odot}$, and 
the factor $\omega_F^2 \equiv \Omega_F(\Omega_h-\Omega_F)/\Omega_h^2$
depends on the angular velocity of the field lines $\Omega_F$ 
relative to that of the hole $\Omega_h$ and is  
taken to have its maximum value of $\omega_F^2 = (0.5)^2$ (e.g. 
Blandford 1990). The magnetic field strength is not expected to 
exceed the Eddington magnetic field strength, given by 
$B_{E} \approx 6 \times 10^4 M_8^{-1/2}$ G 
(e.g. Dermer, Finke, \& Menon 2008). 
Rewriting equation (1) in terms of the black hole spin $j$ and normalized
magnetic field strength $b \equiv B_{p0}/B_E$, we have 
\begin{equation}
jb_{BZ} \approx (B_{E,4}M_8)^{-1} 
\sqrt{(5L_{44})} \approx \sqrt{(17.5 L_j/L_{E})} ,
\end{equation}
where $B_{E,4}$ is the Eddington magnetic field strength in units
of $10^4$ G, $B_{E,4} \approx 6M_8^{-1/2}$, 
$L_{44}$ is the beam power in units of $10^{44}$ erg/s,
and $L_E$ is the Eddington luminosity, 
$L_E \approx 1.26 \times 10^{46} M_8$ erg/s.

Equation (2) allows a determination of the quantity  
$jb_{BZ}$  
when observations allow determinations of the beam power of the 
source, $L_j$, and the black hole mass, $M$.
The parameter $jb_{BZ}$ provides a lower bound on 
$j$ and a lower bound on $b$ in the context of the BZ model 
since the maximum value of $j$ is 
unity and the maximum value of $b$ is unity. When this parameter, 
referred to as the ``j-b'' parameter, 
is equal to one, it suggests that both $j$ and $b$ 
are close to one. If the ``j-b'' parameter exceeds one, it suggests
a problem with the underlying model (in this case, the BZ model). 
Thus, the ``j-b'' parameter indicated by
equation (2) provides a lower bound on $j$ and a lower bound on 
$b$ in the context of the BZ model, and provides a test of the BZ model.  

Models may be characterized by comparison with BZ model predictions
by writing $L_j = \kappa L_j(BZ)$.  Then $jb$ of that model is 
$jb = jb_{BZ}/\sqrt{\kappa}$.  If $jb_{BZ}$ exceeds unity, then 
models with 
values of $\kappa$ with $\kappa \geq [jb_{BZ}]^2$ 
would be indicated.   
Consider, for example, the hybrid model proposed by Meier (1999).  
This model includes characteristics of the
Blandford-Payne (Blandford \& Payne 1982) and Blandford-Znajek 
(Blandford \& Znajek 1977) models. 
The beam power produced in this model is 
$L_j(M) \approx 10^{44} j^2 M_8^2 B_4^2$ 
erg/s (Meier 1999), 
or  $\kappa(M) \simeq 5$, and $jb_{M} \simeq jb_{BZ}/\sqrt{5}$.  

Requiring that $jb$ satisfy $jb \leq 1$ allows constraints to be 
placed on $\kappa$, and this indicates which models can account for the
characteristics of these systems.

\section{RESULTS}
\label{results}

The method is applied to the samples of 19 very powerful FRII 
radio galaxies (referred to as FRIIb sources) 
and 29 central dominant galaxies (CDGs) studied by 
Daly (2009).  The FRIIb sources have radio powers at least
a factor of ten above the classical FRI/FRII transition 
(Fanaroff \& Riley 1974). 
Black hole masses and beam powers are available for all
of these sources; the masses are listed in Tables 1 and 2
of Daly (2009). Beam powers for the powerful classical double radio 
galaxies are obtained from Guerra et al. (2000), Wan, Daly, \& Guerra
(2000), and 
O'Dea et al. (2009); note that these beam powers are independent of
offsets of the extended radio source from minimum energy conditions
(O'Dea et al. 2009).
Beam powers for the 
CDGs are obtained from Rafferty et al. (2006), and also are 
independent of offsets from minimum energy conditions.  
Almost all of the radio sources associated with CDGs 
have FRI radio source structure
or amorphous radio structure, with a few exceptions such as 
Cygnus A (Birzan et al. 2008).  
The 
name, redshift, total beam power, black hole mass, 
and Eddington magnetic field strength 
are listed for each source
in Tables 1 and 2 for the FRIIb and CDG sources, respectively. 
The beam powers and black hole masses were combined to solve 
for $jb_{BZ}$ using equation 
(2) and are listed in Tables 1 and 2.  Values of $jb_{M}$ were
obtained explicitly for the FRIIb sources using $\kappa(M) \approx 5$ 
so $jb_M \approx jb_{BZ}/\sqrt(5)$, 
and are included in Table 1. 
The value of $\sqrt(jb_{BZ})$ is listed in the final column of Tables 
1 and 2, and is discussed below. 
For simplicity, the average values of 
asymmetric error bars was used. 

The parameter $jb_{BZ}$ is shown as a function of black hole mass in Figure 1. 
Analyzing the two samples 
separately, there is no indication of a dependence of $jb_{BZ}$ on black
hole mass (see Figure 1).  Since $jb_{BZ}$, and $jb_{M}$, are proportional to 
$\sqrt{L_j/L_{E}}$ (see eq. 2), this is equivalent to finding no 
dependence of $L_j/L_E$ on black hole mass.  

The parameter $jb_{BZ}$ is shown 
as a function of redshift in Figure 2. Each sample clearly exhibits a 
dependence of $jb$ on redshift (see Figure 2). The sample of 19 powerful
FRIIb sources has the dependence $Log(jb_{BZ}) = 
(0.92 \pm 0.24)~Log(1+z)~-(0.34 \pm 0.06)$.  Given that these are the most
powerful radio sources at each redshift and that they are drawn from a 
complete sample of sources, this represents the envelope of the
distribution of $jb$ values as a function of redshift. Thus, this redshift
dependence can be interpreted as the evolution of the maximum
value of $jb$ as a function of redshift, and is obtained in the context
of the BZ model.  The normalization of $jb$ decreases by a factor of 
about $1/\sqrt(5)$ if the Meier model is considered, but the redshift 
behavior is unaffected.  

The beam power is shown as a function of black hole mass in 
Figure 3.  The CDGs have much lower 
beam powers than the FRIIb sources, though the sources are powered
by black holes with similar mass. 
   
The value of $j$ and $b$ may be independent, or the value of $b$ could
depend upon $j$.  The results of Daly et al. (2009) 
and Daly \& Guerra (2002) suggest that 
$B_{p0} \propto j$, that is, $b \propto j$.  Setting $b = b_* j$, 
and noting that $j \leq 1$ and $b \leq 1$, the
results for $jb$ 
listed in Table 1 indicate that $0.4 \leq b_* \leq 1$ for the
BZ model.  Thus, $b_*$ is of order unity, or $b \approx j$.  In this 
case, $jb \approx j^2$, and the redshift evolution of the 
maximum value of $j$ varies as $Log(j) =(0.46 \pm 0.12)~Log(1+z) - (0.17 \pm 
0.03)$. The values of $j$ obtained using $j \approx \sqrt(jb)$ are listed
in the final column of Tables 1 and 2 and are obtained in the context of the 
BZ  model. 

\section{DISCUSSION}
\label{discussion}

All of the empirically determined $jb_{BZ}$ parameters are consistent with 
$jb_{BZ} \leq 1$ at about the 1 sigma level, thus the current data 
do not require modifications to the BZ model (see Tables 1 and 2 
and Figure 1).  It is interesting that 
many sources have $jb_{BZ} \simeq 1$, indicating that both $j$ and
$b$ are close to one if the BZ model provides an accurate description
of the physics of these systems. 
Almost all of the empirically determined $jb(M)$ parameters are less than
one, indicating that either $j$ or $b$ is less than one if this model
provides an accurate description of these systems. 

As noted in section \ref{method}, the parameter $jb$ provides a lower 
bound on $j$ and a lower bound on $b$ since neither is expected to 
exceed unity.  The lower bounds on $j$ obtained for FRIIb sources 
in the BZ and the 
Meier models are about 0.4 and 0.2, respectively (see Table 1). These
are larger than the lower bound of $j_{min} = 0.12 \pm 0.01$
obtained for these same sources by Daly (2009) 
using a model-independent approach, assuming only that the 
large-scale outflow from the AGN is powered by the spin energy of the 
black hole. 
Combining the results obtained here
with those obtained by Daly (2009) indicates that only a small fraction of the 
spin energy per unit black hole mass, $r$,  
of a system is extracted during one particular outflow event for 
the FRIIb sources studied and most of the CDGs. 
Thus, a single outflow event does not significantly reduce the spin
of the black hole for FRIIb sources and most radio sources associated with 
CDGs. 

The parameter $jb_{BZ}$ is clearly increasing with redshift (see Figure 2),
and is roughly $\propto (1+z)^{0.92 \pm 0.24}$ for the FRIIb sources. 
The fact that lower $jb_{BZ}$ sources are missing at high redshift results
from the radio power selection effect, since all of the higher redshift sources
are from the 3CRR catalog (Laing, Riley, \& Longair 1983); obviously, 
this does not explain the lack of high $jb$ sources at low redshift. 
Given that the sources with high radio power
have high beam power, and, for a fixed black hole mass, 
sources with high beam power have high spin
(see eq. 1), the FRIIb sources studied here are likely to 
have spin values that are close to the maximal values for sources
at that redshift. Thus, the evolution of spin of these systems
should be considered to be the envelope of the distribution, presenting
the maximal spin as a function of redshift. That is, 
the black holes at the centers of massive elliptical
galaxies that produce FRIIb radio sources 
have maximal spin values that are slowly decreasing as 
the universe evolves. 

The results of Daly (2009) suggest that the energy extracted per
unit black hole mass, $r$,  during
each outflow event is independent of source redshift and black hole mass,
and is roughly constant for all of the FRIIb sources studied. 
Given the selection effects affecting the CDGs, the data were consistent
with those sources having a value of $j_{min}$ similar to that of the FRIIb 
sources.  
The lower bounds on $j$ obtained here in the context of the BZ model 
are consistent with $j_{min}$ of $0.12 \pm 0.01$ obtained 
for the FRIIb sources by Daly (2009) 
and with most of the CDGs at about 1 to 2 $\sigma$.  

In the case that $b \simeq j$, as discussed in section \ref{results} and
as indicated by detailed studies of FRIIb sources (e.g. Daly et al. 2009), 
the value of $j$ is obtained by taking the square root of the values of 
$jb$ listed in Tables 1 and 2, and is listed in the final columns of
each table.  In this case, all of the FRIIb and CDG sources have values of 
$j$ that are consistent with the ``model-independent'' minimum value 
$j_{min} = 0.12 \pm 0.01$ obtained by Daly (2009). 
In addition, this implies that 
the maximum value of $j$ evolves as 
$Log(j) =  (0.46 \pm 0.12)~Log(1+z) - (0.17 \pm 
0.03)$.   The evolution of the spins of supermassive black holes 
may be used to study the accretion and merger history of these sources, 
as discussed, for example, by King \& Pringle (2006, 2007), 
Berti \& Volonteri (2008), and 
King, Pringle, and Hofmann (2008). 
It is particularly intriguing that the redshift evolution obtained
here for FRIIb sources is very similar to that 
predicted by the detailed model of King, Pringle, 
\& Hofmann (2008) for spin down by gas accretion through a 
series of repeated accretion episodes. In comparing the results obtained
here with theoretical predictions, it should be kept in mind that all
of the CDGs are by definition associated with giant elliptical galaxies,
and the FRIIb sources are likely to evolve into CDGs 
(Lilly \& Longair 1984; Best et al. 1998;
McLure et al. 2004). Due to the way each sample is selected, the 
CDGs allow a glimpse into the spin values of sources with low radio 
power and low beam power, while the FRIIb sources are the most
powerful extended radio sources at their respective redshifts.

The combination of the results obtained here with those obtained
earlier suggest that each source may undergo multiple outflow
events.  The energy extracted per outflow event is roughly
constant, suggesting very similar physical conditions in the source
when the outflow event is triggered, and the  
energy extracted per outflow event is a small fraction of the total
available energy for most of the sources 
(Daly 2009). Thus, each source may undergo numerous outflow
events. These outflow events could significantly affect the 
gaseous medium in the vicinity of the radio source.  This heating 
of the gaseous environment of the source 
could cause the radio source structure to shift from 
FRII to FRI structure. This is consistent with the ideas that 
very powerful 3CRR radio galaxies evolve into CDGs (Lilly \& Longair 1984;
Best et al. 1998; McLure et al. 2004), that 
different radio structures are determined by the source environment 
(e.g. Hardee et al. 1992; O'Donoghue, Eilek, \& Owen 1993; 
Burns et al. 1994; Bicknell 1995; Barai \& Wiita 2007), 
and that the evolution of the
gaseous environments could be due in part to the large-scale outflows 
(e.g. Silk \& Rees 1998; Eilek \& Owen 2002).  The CDGs do have much lower
beam powers than the FRIIb sources, which may also explain the 
differences in the radio source structures. 

\section{Summary}

Both the CDG and FRIIb samples are illustrative of outflows from massive black 
holes associated with giant elliptical galaxies since FRIIb sources
are expected to evolve into CDGs
(Lilly \& Longair 1984; Best et al. 1998;
McLure et al. 2004).  The CDGs
allow a study of black hole spins and magnetic field strengths of
lower power radio sources that have a range of beam power, black hole
spin, and magnetic field strengths.  The FRIIb sources allow a study of 
black hole spins and magnetic field strengths of the most powerful sources
with large-scale outflows at their respective redshifts; these are the 
sources that have the largest radio power, beam power, and black hole spin. 

These sources were studied in the context of the BZ and Meier models,
which allowed the beam power and black hole mass of each source to be 
combined to obtain lower bounds on the black hole spin and strength of the 
magnetic field close to the black hole.  The lower bounds on the black hole
spin obtained for FRIIb sources were significantly larger than those
obtained by Daly (2009) by studying the outflow energy and black hole mass
of each system.  This implies that, if the black hole spin powers the 
large-scale outflows in these systems, 
then the energy extracted in each outflow
event is a small fraction of the total spin energy available. Thus, 
the evolution of the spin with redshift is a reflection of the 
merger and accretion history of the source, and is not significantly 
affected by energy losses associated with the outflow. And, each source
may undergo numerous outflow events. 

Independent empirical work has shown that the magnetic field 
strength is proportional to the black hole spin 
(e.g. Daly et al. 2009).  This means that the
values of $jb$ obtained here can be used to obtain estimates of the 
black hole spin $j$; the normalization between $b$ and $j$ 
is of order unity, but is not
precisely known. Values of $j$ were obtained in the context of the BZ
model for the FRIIb and CDGs.  The values obtained for FRIIb sources range
from about 0.7 to 1, while those for CDGs range from about 0.14 to 0.8. 
If considered in the context of the Meier (1999) model, these values 
decrease by a factor of about 1.5. 
All of the values are greater than or similar to the 
``model-independent'' minimum value
of $j_{min} \approx 
0.12 \pm 0.01$ obtained by Daly (2009). Losses due to outflows
will affect the spin of the black hole when that spin is close to 
$j_{min}$.  Thus, this could affect the spins of black holes of 
CDGs with low
values of $j$, but is unlikely to affect the those 
associated with FRIIb sources.

\acknowledgements
It is a pleasure to thank the referee for very helpful comments
and suggestions on this work. 
This work was supported in part by U. S. National Science
Foundation grant AST-0507465.

\begin{deluxetable}{lccclllc}
\tablewidth{0pt}
\tablecaption{FRIIb Black Hole Properties}
\tablehead{
\colhead{Source} &\colhead{z} & \colhead{$L_j$ } &  \colhead{$M$} 
&\colhead{$B_{E}$} &\colhead{$jb$} &\colhead{$jb$}
&\colhead{$j$}\\
&&\colhead{$(10^{44}$ erg/s)} &\colhead{$(10^8 M_{\odot}$)}
&\colhead{$(10^{4}$ G)} &\colhead{(BZ)} &\colhead{(M)} &\colhead{(BZ)}
} 
\startdata  					
3C 405 	& 	0.056 	&$ 	47 	\pm 	8 	$&$ 	25 	\pm 	7 	$&$ 	1.2 	\pm 	0.2 	$&$ 	0.51 	\pm 	0.08 	$&$ 	0.23 	\pm 	0.04 	$&$ 	0.72 	\pm 	0.06 	$\\
3C 244.1 	& 	0.43 	&$ 	14 	\pm 	4 	$&$ 	9.5 	\pm 	6.6 	$&$ 	1.9 	\pm 	0.7 	$&$ 	0.44 	\pm 	0.16 	$&$ 	0.2 	\pm 	0.07 	$&$ 	0.67 	\pm 	0.12 	$\\
3C 172 	& 	0.519 	&$ 	31 	\pm 	8 	$&$ 	7.8 	\pm 	5.4 	$&$ 	2.2 	\pm 	0.7 	$&$ 	0.74 	\pm 	0.27 	$&$ 	0.33 	\pm 	0.12 	$&$ 	0.86 	\pm 	0.16 	$\\
3C 330 	& 	0.549 	&$ 	80 	\pm 	20 	$&$ 	13 	\pm 	9 	$&$ 	1.7 	\pm 	0.6 	$&$ 	0.93 	\pm 	0.34 	$&$ 	0.41 	\pm 	0.15 	$&$ 	0.96 	\pm 	0.18 	$\\
3C 427.1 	& 	0.572 	&$ 	31 	\pm 	8 	$&$ 	14 	\pm 	10 	$&$ 	1.6 	\pm 	0.5 	$&$ 	0.55 	\pm 	0.2 	$&$ 	0.25 	\pm 	0.09 	$&$ 	0.74 	\pm 	0.14 	$\\
3C 337 	& 	0.63 	&$ 	20 	\pm 	6 	$&$ 	9.1 	\pm 	6.2 	$&$ 	2 	\pm 	0.7 	$&$ 	0.56 	\pm 	0.2 	$&$ 	0.25 	\pm 	0.09 	$&$ 	0.75 	\pm 	0.14 	$\\
3C34 	& 	0.69 	&$ 	65 	\pm 	9 	$&$ 	16 	\pm 	11 	$&$ 	1.5 	\pm 	0.5 	$&$ 	0.74 	\pm 	0.27 	$&$ 	0.33 	\pm 	0.12 	$&$ 	0.86 	\pm 	0.15 	$\\
3C441 	& 	0.707 	&$ 	65 	\pm 	12 	$&$ 	18 	\pm 	12 	$&$ 	1.4 	\pm 	0.5 	$&$ 	0.72 	\pm 	0.26 	$&$ 	0.32 	\pm 	0.12 	$&$ 	0.85 	\pm 	0.15 	$\\
3C 55 	& 	0.72 	&$ 	180 	\pm 	50 	$&$ 	14 	\pm 	10 	$&$ 	1.6 	\pm 	0.6 	$&$ 	1.3 	\pm 	0.5 	$&$ 	0.58 	\pm 	0.22 	$&$ 	1.14 	\pm 	0.22 	$\\
3C 247 	& 	0.749 	&$ 	35 	\pm 	9 	$&$ 	26 	\pm 	18 	$&$ 	1.2 	\pm 	0.4 	$&$ 	0.43 	\pm 	0.17 	$&$ 	0.19 	\pm 	0.07 	$&$ 	0.66 	\pm 	0.13 	$\\
3C 289 	& 	0.967 	&$ 	85 	\pm 	19 	$&$ 	27 	\pm 	21 	$&$ 	1.2 	\pm 	0.4 	$&$ 	0.66 	\pm 	0.27 	$&$ 	0.3 	\pm 	0.12 	$&$ 	0.82 	\pm 	0.16 	$\\
3C 280 	& 	0.996 	&$ 	53 	\pm 	15 	$&$ 	27 	\pm 	21 	$&$ 	1.2 	\pm 	0.4 	$&$ 	0.52 	\pm 	0.22 	$&$ 	0.23 	\pm 	0.1 	$&$ 	0.72 	\pm 	0.15 	$\\
3C 356 	& 	1.079 	&$ 	250 	\pm 	85 	$&$ 	28 	\pm 	22 	$&$ 	1.1 	\pm 	0.5 	$&$ 	1.12 	\pm 	0.49 	$&$ 	0.5 	\pm 	0.22 	$&$ 	1.06 	\pm 	0.23 	$\\
3C 267 	& 	1.144 	&$ 	190 	\pm 	50 	$&$ 	24 	\pm 	20 	$&$ 	1.2 	\pm 	0.5 	$&$ 	1.04 	\pm 	0.45 	$&$ 	0.47 	\pm 	0.2 	$&$ 	1.02 	\pm 	0.22 	$\\
3C 324 	& 	1.21 	&$ 	150 	\pm 	55 	$&$ 	37 	\pm 	30 	$&$ 	1 	\pm 	0.4 	$&$ 	0.76 	\pm 	0.35 	$&$ 	0.34 	\pm 	0.15 	$&$ 	0.87 	\pm 	0.20 	$\\
3C 437 	& 	1.48 	&$ 	710 	\pm 	180 	$&$ 	24 	\pm 	22 	$&$ 	1.2 	\pm 	0.5 	$&$ 	2.03 	\pm 	0.95 	$&$ 	0.91 	\pm 	0.42 	$&$ 	1.42 	\pm 	0.33 	$\\
3C 68.2 	& 	1.575 	&$ 	210 	\pm 	68 	$&$ 	35 	\pm 	32 	$&$ 	1 	\pm 	0.5 	$&$ 	0.91 	\pm 	0.45 	$&$ 	0.41 	\pm 	0.2 	$&$ 	0.96 	\pm 	0.23 	$\\
3C 322 	& 	1.681 	&$ 	510 	\pm 	140 	$&$ 	32 	\pm 	30 	$&$ 	1.1 	\pm 	0.5 	$&$ 	1.48 	\pm 	0.73 	$&$ 	0.66 	\pm 	0.33 	$&$ 	1.22 	\pm 	0.30 	$\\
3C 239 	& 	1.79 	&$ 	480 	\pm 	170 	$&$ 	37 	\pm 	36 	$&$ 	1 	\pm 	0.5 	$&$ 	1.35 	\pm 	0.69 	$&$ 	0.6 	\pm 	0.31 	$&$ 	1.16 	\pm 	0.30 	$\\
\enddata
\label{Table1}
\end{deluxetable}

\begin{deluxetable}{lccclllc}
\tablewidth{0pt}
\tablecaption{CDG Black Hole Properties}
\tablehead{
\colhead{Source} &\colhead{z} & \colhead{$L_j$ } &  \colhead{$M$} 
&\colhead{$B_{E}$} &\colhead{$jb$} 
&\colhead{$j$}\\
&&\colhead{$(10^{44}$ erg/s)} &\colhead{$(10^8 M_{\odot}$)}
&\colhead{$(10^{4}$ G)} &\colhead{(BZ)} &\colhead{(BZ)} } 
\startdata

        M84          	& 	0.0035 	&$ 	0.01 	\pm 	0.011 	$&$ 	3.4 	\pm 	0.9 	$&$ 	3.2 	\pm 	0.4 	$&$ 	0.02 	\pm 	0.022 	$&$ 	0.14 	\pm 	0.08 	$\\
        M87          	& 	0.0042 	&$ 	0.06 	\pm 	0.026 	$&$ 	8.6 	\pm 	2.9 	$&$ 	2 	\pm 	0.3 	$&$ 	0.031 	\pm 	0.017 	$&$ 	0.18 	\pm 	0.05 	$\\
        Centaurus    	& 	0.011 	&$ 	0.074 	\pm 	0.038 	$&$ 	8.6 	\pm 	2.9 	$&$ 	2 	\pm 	0.3 	$&$ 	0.035 	\pm 	0.021 	$&$ 	0.19 	\pm 	0.06 	$\\
        HCG 62       	& 	0.014 	&$ 	0.039 	\pm 	0.042 	$&$ 	5.7 	\pm 	2.9 	$&$ 	2.5 	\pm 	0.6 	$&$ 	0.031 	\pm 	0.037 	$&$ 	0.18 	\pm 	0.10 	$\\
        A262         	& 	0.016 	&$ 	0.097 	\pm 	0.051 	$&$ 	8.6 	\pm 	2.9 	$&$ 	2 	\pm 	0.3 	$&$ 	0.04 	\pm 	0.025 	$&$ 	0.20 	\pm 	0.06 	$\\
        Perseus      	& 	0.018 	&$ 	1.5 	\pm 	0.7 	$&$ 	17 	\pm 	7 	$&$ 	1.4 	\pm 	0.3 	$&$ 	0.11 	\pm 	0.07 	$&$ 	0.33 	\pm 	0.10 	$\\
        PKS 1404-267 	& 	0.022 	&$ 	0.2 	\pm 	0.18 	$&$ 	5.7 	\pm 	2.9 	$&$ 	2.5 	\pm 	0.6 	$&$ 	0.07 	\pm 	0.07 	$&$ 	0.26 	\pm 	0.13 	$\\
        A2199        	& 	0.03 	&$ 	2.7 	\pm 	1.6 	$&$ 	20 	\pm 	9 	$&$ 	1.3 	\pm 	0.3 	$&$ 	0.14 	\pm 	0.1 	$&$ 	0.37 	\pm 	0.13 	$\\
        A2052        	& 	0.035 	&$ 	1.5 	\pm 	1.4 	$&$ 	17 	\pm 	7 	$&$ 	1.4 	\pm 	0.3 	$&$ 	0.11 	\pm 	0.11 	$&$ 	0.33 	\pm 	0.16 	$\\
        2A 0335+096  	& 	0.035 	&$ 	0.24 	\pm 	0.15 	$&$ 	14 	\pm 	7 	$&$ 	1.6 	\pm 	0.4 	$&$ 	0.048 	\pm 	0.038 	$&$ 	0.22 	\pm 	0.09 	$\\
        MKW 3S       	& 	0.045 	&$ 	4.1 	\pm 	2.3 	$&$ 	8.6 	\pm 	2.9 	$&$ 	2 	\pm 	0.3 	$&$ 	0.26 	\pm 	0.17 	$&$ 	0.51 	\pm 	0.17 	$\\
        A4059        	& 	0.048 	&$ 	0.96 	\pm 	0.62 	$&$ 	29 	\pm 	14 	$&$ 	1.1 	\pm 	0.3 	$&$ 	0.068 	\pm 	0.056 	$&$ 	0.26 	\pm 	0.11 	$\\
        Hydra A      	& 	0.055 	&$ 	4.3 	\pm 	1.3 	$&$ 	11 	\pm 	4 	$&$ 	1.8 	\pm 	0.3 	$&$ 	0.23 	\pm 	0.11 	$&$ 	0.48 	\pm 	0.11 	$\\
        A85          	& 	0.055 	&$ 	0.37 	\pm 	0.24 	$&$ 	29 	\pm 	14 	$&$ 	1.1 	\pm 	0.3 	$&$ 	0.042 	\pm 	0.035 	$&$ 	0.21 	\pm 	0.08 	$\\
        Cygnus A     	& 	0.056 	&$ 	13 	\pm 	7 	$&$ 	29 	\pm 	14 	$&$ 	1.1 	\pm 	0.3 	$&$ 	0.25 	\pm 	0.18 	$&$ 	0.50 	\pm 	0.18 	$\\
        Sersic 159/03 	& 	0.058 	&$ 	7.8 	\pm 	5.4 	$&$ 	17 	\pm 	9 	$&$ 	1.4 	\pm 	0.4 	$&$ 	0.25 	\pm 	0.21 	$&$ 	0.50 	\pm 	0.21 	$\\
        A133         	& 	0.06 	&$ 	6.2 	\pm 	1.4 	$&$ 	20 	\pm 	10 	$&$ 	1.3 	\pm 	0.3 	$&$ 	0.21 	\pm 	0.11 	$&$ 	0.46 	\pm 	0.13 	$\\
        A1795        	& 	0.063 	&$ 	1.6 	\pm 	1.4 	$&$ 	23 	\pm 	11 	$&$ 	1.3 	\pm 	0.3 	$&$ 	0.099 	\pm 	0.099 	$&$ 	0.31 	\pm 	0.16 	$\\
        A2029        	& 	0.077 	&$ 	0.87 	\pm 	0.27 	$&$ 	60 	\pm 	36 	$&$ 	0.77 	\pm 	0.23 	$&$ 	0.045 	\pm 	0.03 	$&$ 	0.21 	\pm 	0.07 	$\\
        A478         	& 	0.081 	&$ 	1 	\pm 	0.5 	$&$ 	26 	\pm 	14 	$&$ 	1.2 	\pm 	0.3 	$&$ 	0.073 	\pm 	0.055 	$&$ 	0.27 	\pm 	0.10 	$\\
        A2597        	& 	0.085 	&$ 	0.67 	\pm 	0.58 	$&$ 	8.6 	\pm 	2.9 	$&$ 	2 	\pm 	0.3 	$&$ 	0.1 	\pm 	0.1 	$&$ 	0.32 	\pm 	0.15 	$\\
        3C 388       	& 	0.092 	&$ 	2 	\pm 	1.8 	$&$ 	17 	\pm 	7 	$&$ 	1.4 	\pm 	0.3 	$&$ 	0.13 	\pm 	0.13 	$&$ 	0.36 	\pm 	0.18 	$\\
        PKS 0745-191 	& 	0.103 	&$ 	17 	\pm 	9 	$&$ 	31 	\pm 	16 	$&$ 	1.1 	\pm 	0.3 	$&$ 	0.27 	\pm 	0.19 	$&$ 	0.52 	\pm 	0.19 	$\\
        Hercules A   	& 	0.154 	&$ 	3.1 	\pm 	2.5 	$&$ 	20 	\pm 	11 	$&$ 	1.3 	\pm 	0.4 	$&$ 	0.15 	\pm 	0.14 	$&$ 	0.38 	\pm 	0.19 	$\\
        Zw 2701      	& 	0.214 	&$ 	60 	\pm 	62 	$&$ 	17 	\pm 	9 	$&$ 	1.4 	\pm 	0.4 	$&$ 	0.7 	\pm 	0.8 	$&$ 	0.84 	\pm 	0.48 	$\\
        MS 0735.6+7421 	& 	0.216 	&$ 	69 	\pm 	51 	$&$ 	20 	\pm 	11 	$&$ 	1.3 	\pm 	0.4 	$&$ 	0.69 	\pm 	0.65 	$&$ 	0.83 	\pm 	0.39 	$\\
        4C 55.16     	& 	0.242 	&$ 	4.2 	\pm 	3 	$&$ 	14 	\pm 	7 	$&$ 	1.6 	\pm 	0.4 	$&$ 	0.2 	\pm 	0.18 	$&$ 	0.45 	\pm 	0.20 	$\\
        A1835        	& 	0.253 	&$ 	18 	\pm 	13 	$&$ 	54 	\pm 	36 	$&$ 	0.81 	\pm 	0.27 	$&$ 	0.21 	\pm 	0.21 	$&$ 	0.46 	\pm 	0.22 	$\\
        Zw 3146      	& 	0.291 	&$ 	58 	\pm 	42 	$&$ 	74 	\pm 	53 	$&$ 	0.7 	\pm 	0.25 	$&$ 	0.33 	\pm 	0.33 	$&$ 	0.57 	\pm 	0.29 	$\\
\enddata
\label{Table}
\end{deluxetable}

\begin{figure}
    \centering
    \includegraphics[width=\textwidth]{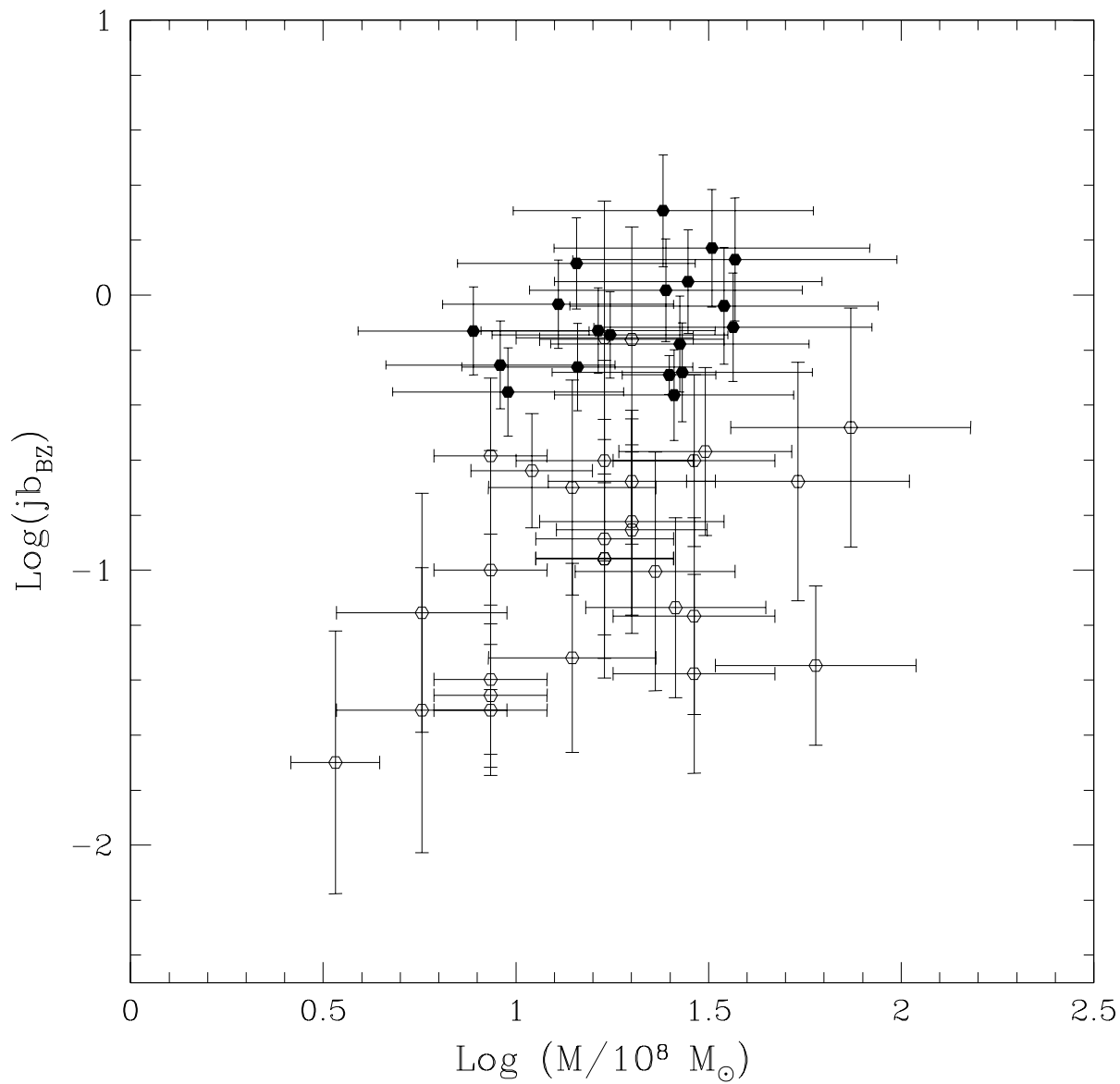}
       \caption{Distribution of 'j-b' parameters obtained in 
the context of the BZ model as a function of 
black hole mass. 
The 19 sources associated 
with very powerful classical double 
radio galaxies are indicated by solid 
circles, and the 29 sources associated with 
CDGs are indicated by open circles. 
One source, Cygnus A (3C 405) is included
in both samples.  }
          \label{figjBofM}
    \end{figure}
\begin{figure}
    \centering
    \includegraphics[width=\textwidth]{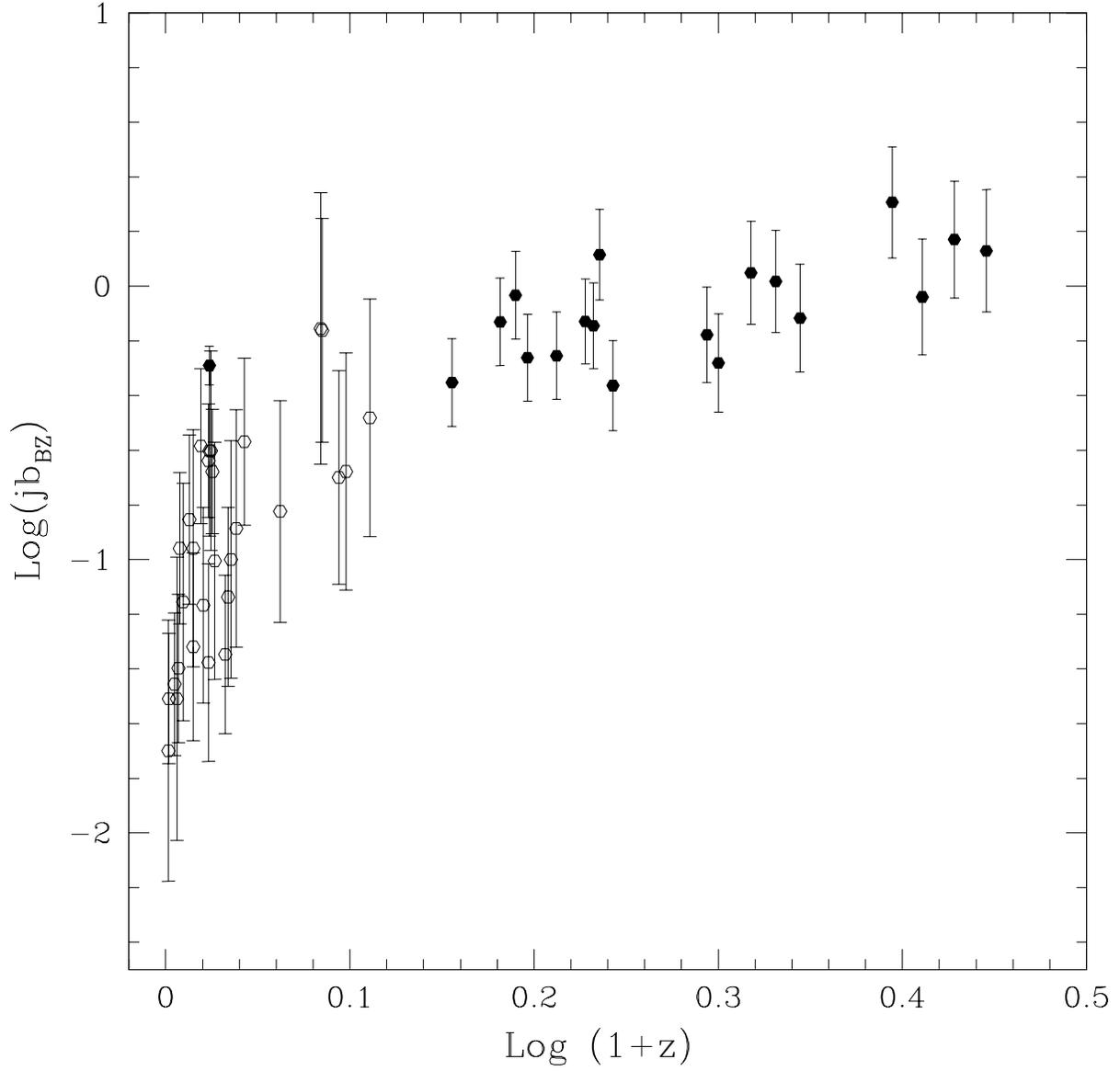}
       \caption{Distribution of 'j-b' parameters obtained in the 
context of the BZ model as a function of 
redshift. The symbols are as in Fig. \ref{figjBofM}.  }
          \label{figjBofz}
    \end{figure}

\begin{figure}
    \centering
    \includegraphics[width=\textwidth]{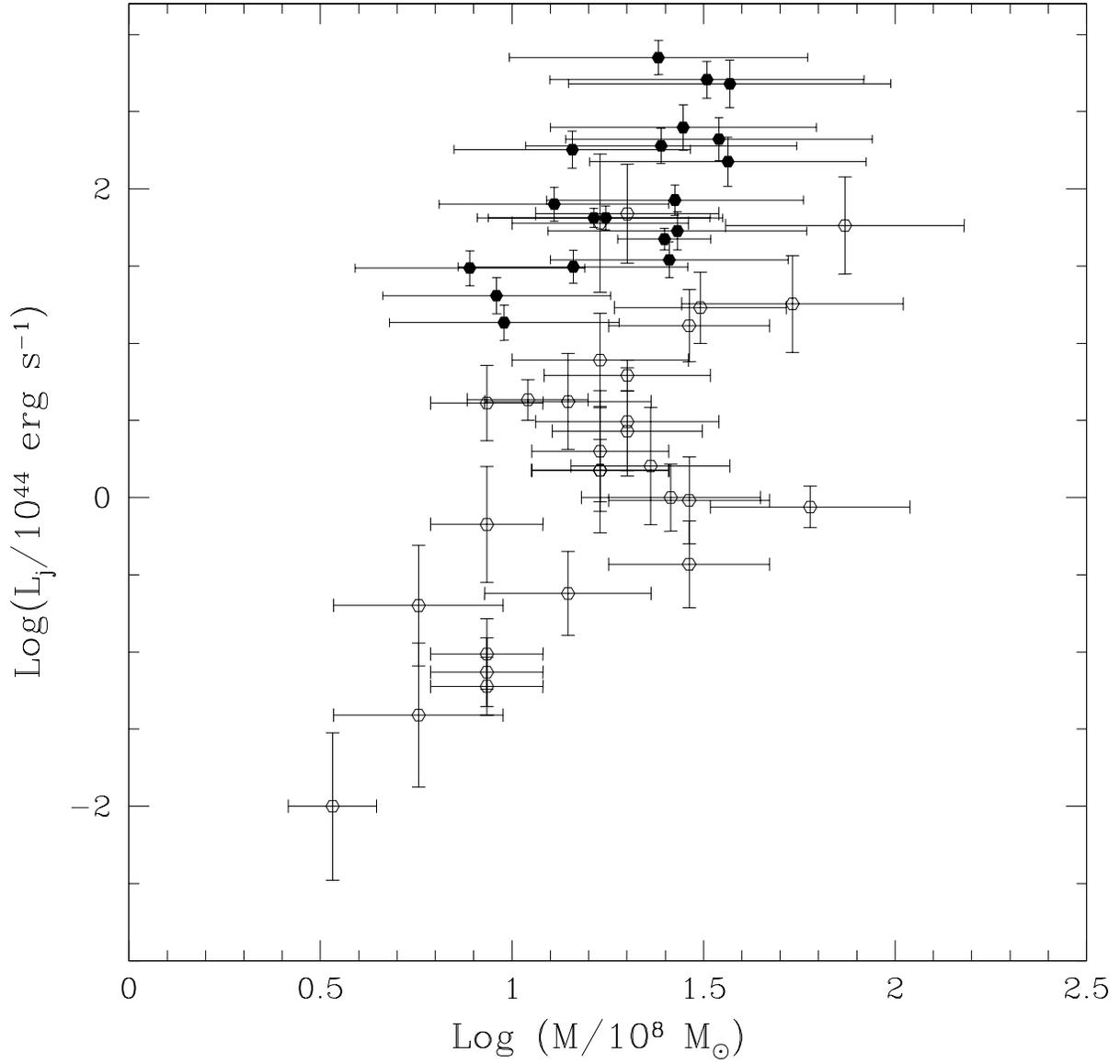}
       \caption{Beam powers as a function of black hole mass.
The symbols are as in Fig. \ref{figjBofM}. }
          \label{figLofM}
    \end{figure}

\end{document}